\documentstyle[11pt]{article}
\renewcommand{\baselinestretch}{1.0}
\def\ds{\displaystyle}
\newcommand{\bit}{\begin{itemize}}
\newcommand{\eit}{\end{itemize}}
\newcommand{\be}{\begin{equation}}
\newcommand{\ee}{\end{equation}}
\begin{document}
\begin{center}
{\huge Discrete symmetries in the Weyl expansion}\break
{\huge for quantum billiards}

\vspace{1.0cm}

{\Large Nicolas Pavloff}
\end{center}

\vspace{0.5 cm}

\noindent Division de Physique Th\'eorique\footnote { Unit\'e de
Recherche des Universit\'es Paris XI et Paris VI associ\'ee au CNRS.},
Institut de Physique Nucl\'eaire, F-91406 Orsay Cedex, France \break

\vspace{0.5 cm}
\begin{center}
{\bf abstract}
\end{center}

	We consider 2 and 3 dimensional quantum billiards with discrete
symmetries. The boundary condition is either Dirichlet or Neumann. We derive
the first terms of the Weyl expansion for the level density projected onto the
irreducible representations of the symmetry group. The formulae require only
the knowledge of the character table of the group and the geometrical
properties (such as surface, perimeter etc ...) of sub-parts of the billiard
invariant under a group transformation. As an illustration, the method is
applied to the icosahedral billiard.

\vspace{5cm}
\noindent PACS numbers :\hfill\break
\noindent 03.65.Sq Semiclassical theories and applications.\hfill\break

\noindent IPNO/TH 94-14 \hspace{0.5cm} {\it submitted to J. Phys.
A: Math. Gen. }\hfill\break
\newpage

	Billiards have been extensively studied in the context of quantum
chaos, both from the semi-classical point of view (periodic orbit theory, see
{\it e.g.} Ref. \cite{Gu90}) and using random matrix theory (for a review see
\cite{Ori91}). In these two approaches it is important to subtract the smooth
part of the level density in order to study the oscillations (trace formula) or
the short range fluctuations (random matrices) around the average behaviour.
The asymptotic behaviour of the smooth part of the level density ---
characterized by the Weyl expansion --- is also of interest in a great variety
of other physical and mathematical problems (see Refs.
\cite{Kac66,See69,BB70,Bal76,Dur82,Ber94}).

	In this paper we study two and three dimensional billiards : a
quantum particle is enclosed in a compact region ${\cal B}$ of space and we
impose to its wave-function $\psi$ Dirichlet or Neumann boundary conditions on
the frontier $\partial {\cal B}$ ($\partial {\cal B}$ may or may not be a
smooth manifold). Thus $\psi$ verifies the following equation :

\begin{eqnarray}\label{e1}
(\Delta + k^2)\psi & = & 0 \qquad \hbox{inside} \quad {\cal B} \; , \\
\psi & = & 0 \qquad \hbox{on} \quad \partial{\cal B} \qquad \hbox{(Dirichlet)}
\nonumber \\
{\bf n . \nabla} \psi & = & 0 \qquad \hbox{on} \quad
\partial{\cal B} \qquad \hbox{(Neumann)} \nonumber \; .
\end{eqnarray}

	We will assume that ${\cal B}$ remains invariant under a discrete
point group ${\cal G}$. Then the eigenfunctions $\{\psi_n\}_{n\geq 0}$ and
eigenvalues $\{k_n\}_{n\geq 0}$ can be classified according to the different
irreducible representations (irreps) of the group. We label these irreps by an
index $(\alpha)$, each has a dimension $d^\alpha$. Hence the total level
density $\rho(k)$ can be written as :

\be\label{e2a} \rho (k) = \sum_{\alpha = 1}^r \rho^{\alpha}(k) \; , \ee

\noindent where $r$ is the total number of non-equivalent irreps of ${\cal G}$,
and $\rho^\alpha(k)$ is the density of levels belonging to irrep $\alpha$ :

\be\label{e2b} \rho^{\alpha}(k) = d^\alpha \sum_{n=0}^{\infty} \delta
(k-k_n^\alpha) \; .\ee

	$\rho(k)$ (resp. $\rho^\alpha(k)$) can be separated in a smooth function
of $k$, $\bar\rho(k)$ (resp. $\bar\rho^\alpha(k)$) plus an oscillating part
$\widetilde\rho(k)$ (resp. $\widetilde\rho^\alpha(k)$). The study of the
oscillating part $\widetilde\rho^\alpha$ needs a careful treatment and has
been addressed in Ref. \cite{Rob89,Cvi89,Lau91,Str93}. Our aim is to write the
first terms of an explicit Weyl expansion for $\bar\rho^\alpha(k)$ :

\be\label{e2c} \bar\rho^\alpha(k) = \bar\rho^\alpha_0(k) + \bar\rho^\alpha_1(k)
+ \cdots \ee

	In 3D $\bar\rho^\alpha_0(k) $ is a volume term of order $k^2$,
$\bar\rho^\alpha_1$ is a surface term of order $k$ and the next order is a
constant (order zero in $k$) edge term $\bar\rho^\alpha_2$. In the following we
specialize the periodic orbit theory initiated in Ref. \cite{Rob89} to the
simple case of zero length trajectories and treat the boundary conditions by
using the Balian and Bloch method \cite{BB70}. We will derive the formulae in
the case of three dimensional billiards and only state the two dimensional
results.

	The projector onto an irreducible invariant subspace $(\alpha)$ is
written as \cite{Wu85} :

\be\label{e3} P^\alpha = {d^\alpha\over |{\cal G}|} \sum_{g \in {\cal G}}
\chi^\alpha(g) U^{\mbox{\dag}} (g) \; , \ee

\noindent where $|{\cal G}|$ is the order of the group, $\chi^\alpha(g)$ is the
character of group element $g$ in irrep $(\alpha)$ and $U(g)$ is the operator
representing the action of $g$ in Hilbert space. The projected level density is
then (see \cite{Rob89}) :

\be\label{e4} \rho^\alpha(k) = -{2k\over\pi} \; \hbox{Im} \;
\hbox{Tr} \{ P^\alpha G(k+i0^+) \} \; , \ee

\noindent where $G(k+i0^+)$ is the retarded Green function of our problem ; for
simplicity we will drop the argument in the following. According to
the ideas of Balian and Bloch, $G$ is written as the free Green function $G_0$
plus a correcting term :

\be\label{e5} G = G_0 + G_1 \; . \ee

	We will here only need the explicit expression of $G_1$ for short
trajectories near the boundary $\partial {\cal B}$. In this case it can be
approximated using the method of images :

\be\label{e6} <{\bf r}|G_1|{\bf r^\prime}> = \epsilon
<{\bf r}|G_0|{\bf r^\prime}_1>  \; ,\ee

\noindent where $\epsilon = \pm 1$ for Dirichlet or Neumann boundary conditions
and ${\bf r^\prime}_1$ is the symmetric of point ${\bf r^\prime}$ with respect
to the plane tangent to $\partial{\cal B}$. Note that (\ref{e6}) is properly
defined only if ${\bf r}$ and ${\bf r^\prime}$ are close to each other and
close to the boundary. (\ref{e6}) is the leading correction (in the $k$
expansion) to $G_0$ arising near the boundary. A more systematic expansion can
be found in Ref. \cite{BB70}. In the Balian--Bloch terminology, we only
consider the first term of the multiple scattering expansion and furthermore
restrict ourselves to zero length orbits (or high wave vector).

	In three dimensions, $G_0$ contributes to the total level density with
a volume term $\bar\rho_0(k)=V k^2/4\pi^2$ ($V$ being the volume of the
billiard), $G_1$ gives the surface term $\bar\rho_1(k)$ and the next order
$\bar\rho_2$ is obtained by pursuing the multiple scattering expansion. We
shall not consider higher orders in the expansion. Surprisingly enough, when
computing the symmetry-projected terms ($\bar\rho^\alpha_0$,
$\bar\rho^\alpha_1$ and $\bar\rho^\alpha_2$), we need not know the form of the
Green function to a better approximation than (\ref{e5},\ref{e6}).

	The contribution of $G_0$ to the level density (\ref{e4}) is (using
(\ref{e3}))

\be\label{e7} -{2k\over\pi} {d^\alpha\over |{\cal G}|} \sum_{g\in {\cal G}}
\chi^\alpha(g) \; \mbox{Im} \int\!\! d^3r
<g{\bf r} | G_0|{\bf r}> \; , \ee

\noindent and the non-oscillating part of this expression is given by the quasi
zero length trajectories going from ${\bf r}$ to $g{\bf r}$. In 3D the elements
of a point group ${\cal G}$ can be either symmetries with respect to a plane,
rotations or improper rotations ({\it ie} product of elements of the two first
types) ; and orbits going from ${\bf r}$ to $g{\bf r}$ do not all contribute at
the same order in the Weyl expansion :
\bit
\item[({\it i})] If $g=e$ is the identity, then $e{\bf r}={\bf r}$ for all
points in ${\cal B}$ and (\ref{e7}) gives the projected volume term
$\bar\rho^\alpha_0(k)$.
\item[({\it ii})] If $g$ is a symmetry with respect to a plane, it leaves
invariant a surface of area $S_g$ (intersection of the plane of symmetry with
${\cal B}$) and then the corresponding term in (\ref{e7}) will contribute to
$\bar\rho^\alpha_1$.
\item[({\it iii})] If $g$ is a rotation, its invariant points in ${\cal B}$ are
located on a segment of length $L_g$ and we will have an edge contribution to
$\bar\rho^\alpha_2$.
\eit

	Terms arising from improper rotations in (\ref{e7}) do not contribute
at order $\bar\rho^\alpha_2$ and we will not consider them here. From
(\ref{e7}) we can readily write the symmetry projected volume term (the first
term in the Weyl expansion) :

\be\label{e8} \bar\rho^\alpha_0 (k) = {(d^\alpha)^2\over |{\cal G}|}
\bar\rho_0 (k) \; . \ee

	This formula is quite general and holds also for smooth potentials. A
similar expression can be derived for continuous symmetry groups as well : the
discrete sum in (\ref{e3}) is simply replaced by an integration on the
invariant measure of the group. To fix the ideas, if we consider a system with
5 fold and 1 fold degenerate levels, formula (\ref{e8}) implies that roughly
speaking, the 5 fold degenerate levels will contribute 25 times more to the
level density than the singly degenerate levels (due to the coefficient
$(d^\alpha)^2$ in (\ref{e8})). This contribution comprises a trivial factor 5
coming from the degeneracy of the levels. But the factor 5 remaining
($25=5\times 5$) is more surprising : when choosing a level at random in the
spectrum, the probability is approximatively 5 times bigger for drawing a 5
fold degenerate level than for a singly degenerate level.

	The contribution of $G_1$ to the level density (\ref{e4}) is of the
same type as Eq. (\ref{e7}). There similarly, the identity, symmetries with
respect to a plane and rotations do not contribute at the same order. The
identity gives a surface contribution $(d^\alpha)^2\bar\rho_1/|{\cal G}|$ (as
for the volume term (\ref{e8}) arising from $g=e$ in (\ref{e7})). Symmetries
contribute to $\bar\rho^\alpha_2$ and we will not consider the effects of
rotations (contributing at higher order). The term following $G_1$ as given by
(\ref{e6}) in the multiple reflexion expansion contributes to the total level
density at order $\bar\rho_2$. Its contribution will be weighted by a factor
$(d^\alpha)^2/|{\cal G}|$ when projected onto a given irrep (this again comes
from $g=e$ in (\ref{e3}) while other elements of the group contribute at higher
orders).

	Gathering all the contributions we get for the surface and edge terms :

\begin{eqnarray}
\bar\rho_1^\alpha(k)={(d^\alpha)^2\over |{\cal G}|} \bar\rho_1(k) & - &
{2k\over\pi} {d^\alpha\over |{\cal G}|} \sum_{g \in {\cal G}sym.}
\chi^\alpha(g) \;
\hbox{Im} \int\!\! d^3r <g{\bf r} | G_0|{\bf r}> \; , \label{e9}\\
\bar\rho_2^\alpha(k)={(d^\alpha)^2\over |{\cal G}|} \bar\rho_2(k) & - &
{2k\over\pi} {d^\alpha\over |{\cal G}|} \sum_{g \in {\cal G}rot.}
\chi^\alpha(g) \;
\hbox{Im} \int\!\! d^3r <g{\bf r} | G_0|{\bf r}> \nonumber \\
 & - & {2 k\over\pi} {d^\alpha\over |{\cal G}|}
\sum_{g \in {\cal G}sym.} \chi^\alpha(g) \;
\hbox{Im} \int\!\! d^3r <g{\bf r} | G_1|{\bf r}> \; . \label{e10}
\end{eqnarray}

	In (\ref{e9},\ref{e10}) the integration must be restricted to
quasi-zero length trajectories. $\sum_{g \in {\cal G}rot.}$ indicates a sum
restricted to the rotations of the group (the same convention holds for
$\sum_{g \in {\cal G}sym.}$). To be specific we will compute one of the
integrals in (\ref{e9},\ref{e10}). Let us consider for instance the last term
of the r.h.s. of Eq. (\ref{e9}). The integral $\int\!\!d^3r$ can be separated
in
a surface integral $\int\!\! dx dy$ along the invariant symmetry plane plus a
term $\int\!\! dz$ along the normal to this plane. Then the modulus $||g{\bf
r}-{\bf r}||=2|z|$ and the free Green function reads (see {\it e.g.} Ref.
\cite{Eco83})

\be\label{e11}
<g{\bf r}|G_0|{\bf r}> = -{\ds e^{\ds ik||g{\bf r}-{\bf r}||}\over \ds 4\pi
||g{\bf r}-{\bf r}||} = -{\ds e^{\ds 2ik|z|}\over \ds 8\pi |z|} \; .\ee

	Hence the integral over $dx dy$ will give a factor $S_g$ (the area of
the invariant surface) and the imaginary part of the integral over $dz$ will
give

\be\label{e12} -{1\over 8\pi} \int_{-\infty}^{+\infty}dz {\sin(2k|z|)\over
|z|}= -{1\over 8} \; . \ee

	Thus the total surface projected term (\ref{e9}) reads :

\be\label{e13}
\bar\rho^\alpha_1(k) = {(d^\alpha)^2\over |{\cal G}|} \bar\rho_1(k)
+ {d^\alpha\over |{\cal G}|} \sum_{g \in {\cal G}sym.} \chi^\alpha(g)
{k S_g\over 4\pi} \; .\ee

	We note that : ({\it i}) if a class of ${\cal G}$ contains one symmetry
it is only formed by symmetries ; ({\it ii}) $\chi^\alpha(g) S_g$ is a class
function, {\it i.e.} it takes the same value for all symmetries belonging to
the same class $C$ (we note $\chi^\alpha(C)S_c$ the value of the class
function). Hence (\ref{e13}) can be written as a sum over those classes of
the group which are formed by symmetries with respect to a plane (noted
$\sum_{Csym.}$ below) :

\be\label{e14}
\bar\rho^\alpha_1(k) = {(d^\alpha)^2\over |{\cal G}|} \bar\rho_1(k)
+ {d^\alpha\over |{\cal G}|} \sum_{Csym.} \chi^\alpha(C) |{\cal G}_c|
{k S_c\over 4\pi} \; .\ee

	In (\ref{e14}) $|{\cal G}_c|$ is the number of elements of the group
belonging to class $C$.

	Similar computations yield for the edge terms :

\be\label{e15}
\bar\rho^\alpha_2(k) = {(d^\alpha)^2\over |{\cal G}|} \bar\rho_2(k)
+ {d^\alpha\over |{\cal G}|} \sum_{Crot.} \chi^\alpha(C) |{\cal G}_c|
{L_c\over 4\pi \sin^2 ( {\ds\beta_c\over \ds 2}) }
+ \epsilon {d^\alpha\over |{\cal G}|} \sum_{Csym.} \chi^\alpha(C) |{\cal G}_c|
{\partial S_c\over 8\pi} \; .\ee

	In (\ref{e15}) $\beta_c$ is the angle of rotation, common (modulo an
irrelevant sign) to all the rotations belonging to a given class $C$ and $L_c$
is the common length of the invariant segments. Here also we use the fact that
if a class contains one rotation, then it contains only rotations.
$\partial S_c$ is the common perimeter of the surfaces invariant under the
symmetries belonging to the same class ({\it i.e.} $\partial S_c$ is the
perimeter of $S_c$).

	Formulae (\ref{e8}), (\ref{e14}) and (\ref{e15}) (together with their
2D counterparts (\ref{e16}) and (\ref{e17})) are the most important results of
this paper. They allow to compute the first terms of the symmetry projected
Weyl expansion knowing only the character table of the group and the surfaces,
lengths and perimeters of sub-parts of ${\cal B}$ invariant under a given class
of group transformations. Note that using elementary group theoretical
properties of the character table (see {\it e.g.} Ref. \cite{Wu85}) one
recovers the total smooth level density when summing the different projected
terms.

	In two dimensions, the present formalism allows to determine the value
of $\bar\rho^\alpha_0(k)$ (which is now a surface term) and of
$\bar\rho^\alpha_1(k)$ (the perimeter correction). The formulae read :

\begin{eqnarray}
\bar\rho_0^\alpha(k) & = & {(d^\alpha)^2\over |{\cal G}|} \bar\rho_0(k)
\label{e16} \; ,\\
\bar\rho_1^\alpha(k) & = & {(d^\alpha)^2\over |{\cal G}|} \bar\rho_1(k) +
{d^\alpha\over |{\cal G}|} \sum_{Csym.} \chi^\alpha(C) |{\cal G}_c|
{L_c\over 2\pi} \label{e17} \; .
\end{eqnarray}

	In (\ref{e16},\ref{e17}) $\bar\rho_0(k)= S k/2\pi$ ($S$ being the area
of the billiard) is the total surface term and $\bar\rho_1(k)= \epsilon L/4\pi$
($L$ is the perimeter) the total perimeter term. $L_c$ is the common length of
the segments invariant under the symmetries belonging to the class considered.
Note that similar results were obtained in Ref. \cite{BTU93} for a smooth
potential (the two-dimensional quartic oscillator) using the Wigner transform
approach.

	An alternative method for deriving the above results would be to work
in reduce configuration space, to find out for each irrep which are the
boundary conditions on the elementary cell and to work out the corresponding
Weyl expansion (as was done in \cite{BGS84} for instance). But for some
symmetry groups the boundary conditions on the elementary cell relevant for
each irrep can be cumbersome (see below the example of the icosahedron) and
also the corresponding Weyl expansion might not be known up to order
$\bar\rho^\alpha_2(k)$ : some irreps lead to mixed Neumann--Dirichlet
conditions or impose phase-shifts when going from one boundary of the
elementary cell to an other (this is the case for the rather simple group
${\cal C}_{3v}$). The method presented here has the advantage of giving simple
and easily applied formulae that do not require detailed knowledge of the
reduced boundary condition.

	As an illustration let us consider a three dimensional icosahedral
billiard with Dirichlet boundary conditions. This billiard was studied in
\cite{Pav93} as a model for faceted metal clusters. The first 565 quantum
levels were determined numerically (counting the degeneracies this leads to
2094 available states). The total symmetry group of the icosahedron is ${\cal
I}_h$ ; to simplify the presentation we will consider here only its subgroup
${\cal I}$ formed by 60 rotations. ${\cal I}$ has 5 classes and 5 non
equivalent irreps, we will present only the results for one of the irreps which
has dimension $d^\alpha=3$ (see table 1). In order to compare the numerical
results with the predictions of Eq. (\ref{e8}), (\ref{e14}) and (\ref{e15}) we
focus on the cumulated level density :

\be\label{e18} \bar N^\alpha(k)=\int_0^k \bar\rho^\alpha(k^\prime)dk^\prime
= \bar N^\alpha_0(k) + \bar N^\alpha_1(k) + \bar N^\alpha_2(k) +
\bar N^\alpha_3(k) + \cdots \ee

	From Ref. \cite{Bal76} we know the total ---{\it i.e.} summed over all
the irreps--- terms $\bar N_0$, $\bar N_1$ and $\bar N_2$ (the explicit
expression is given in \cite{Pav93}). ${\cal I}$ contains no symmetry and we
can apply formulae (\ref{e14},\ref{e15}) if knowing the length of the segments
invariant under rotations (they are listed on table 1). The comparison of the
numerical $N^\alpha$ --- which is a staircase function --- with its smooth
approximation $\bar N^\alpha_0 + \bar N^\alpha_1 + \bar N^\alpha_2$ (from
(\ref{e8},\ref{e14},\ref{e15})) is shown Fig. 1. The difference is also plotted
on this figure : as expected it is of order zero in $k$, and the amplitude of
the oscillations is of order of the degeneracy $d^\alpha=3$. A simple fit gives
the value of the constant next order $\bar N_3^\alpha(k) \simeq 0.288$. The
same agreement is also obtained for the other irreps of ${\cal I}$ or when
studying the irreps of the total group ${\cal I}_h$. It is interesting to note
that the boundary conditions to be applied on the elementary cell when studying
the different irreps of ${\cal I}$ and ${\cal I}_h$ are very difficult to
determine. The method exposed here uses the character table of the group which
comprises the necessary information in an easily accessible form.

\

{\bf Acknowledgements:} It is a pleasure to thank S. C. Creagh and M. Knecht
for fruitful discussions. I am particularly grateful to C. Schmit for his
interest in this study and for the advises he provided at all the stages of
this work.

\newpage

\renewcommand{\baselinestretch}{.7}

\newpage
{\center {\Large {\bf Table captions }}}
\vspace{1cm}

{\bf Table 1.} ~Characters of the classes of ${\cal I}$ for the irrep
considered in the text. For the designation of the classes we use the
conventions of Ref. \cite{Landau}. $L_c$ is for each class (except
for the identity) the length of the invariant segments. $L_c$ is given in
units of the edge length.

\vspace{0.5cm}

\begin{large}
\begin{center}
\begin{tabular} {|c|c|c|c|c|c|} \hline\hline
 & & & & & \\
 Classes & $1 \; C_1= \{ e \}$ & $12 \; C_5$ & $12 \; C^2_5$ & $15 \; C_2$ &
$20 \; C_3$ \\
 & & & & & \\ \hline
 & & & & & \\
 $\chi^\alpha(C)$ & $3$  & ${\ds 1-\sqrt{5}\over \ds 2}$ &
${\ds 1+\sqrt{5}\over \ds 2}$ & $-1$ & $0$ \\
 & & & & & \\ \hline
 & & & & & \\
 $L_c$  & -- & $\sqrt{{\ds 5+\sqrt{\ds 5}\over\ds 2}}$ &
$\sqrt{{\ds 5+\sqrt{\ds 5}\over\ds 2}}$ & ${\ds 1+\sqrt{5}\over \ds 2}$ &
${\ds 2\over\ds \sqrt{3}(3-\sqrt{5})}$ \\
 & & & & & \\ \hline\hline
\end{tabular}
\end{center}
\end{large}

\vspace{1cm}
{\center {\Large {\bf Figure captions }}}
\vspace{1cm}

{\bf Figure 1.} ~bottom part : $N^\alpha(k)$ (staircase function) and its
smooth approximation $\bar N^\alpha_0 + \bar N^\alpha_1 + \bar N^\alpha_2$ as a
function of $k$. $k$ is given in units of the inverse length of the edge of the
icosahedron. Top part : difference of the two quantities $\Delta N^\alpha =
N^\alpha -  (\bar N^\alpha_0 + \bar N^\alpha_1 + \bar N^\alpha_2)$ as a
function of $N^\alpha$. The thin horizontal line is the mean value of $\Delta
N^\alpha$ ($\simeq 0.288$). The bottom plot is limited to the first 30 levels
for legibility. The upper part concerns the first 100 levels ({\it i.e.} the
first 300 quantum states).

\end{document}